\documentclass[11pt,twoside]{article}
\usepackage{asp2014}

\aspSuppressVolSlug
\resetcounters

\begin{document}

\title{ESCAPE - addressing Open Science challenges.}

\author{Mark G. Allen$^1$, Giovanni Lamanna$^2$, Xavier Espinal$^3$, Kay Graf$^4$, Michiel van Haarlem$^5$, Stephen Serjeant$^6$, Ian Bird$^2$, Elena Cuoco$^7$, Jayesh Wagh$^2$ }
\affil{$^1$Observatoire astronomique de Strasbourg, Universit\'{e} de Strasbourg, CNRS, UMR 7550, Strasbourg, F-67000, France, \email{mark.allen@astro.unistra.fr}}
\affil{$^2$Laboratoire d’Annecy de Physique des Particules (LAPP), Universite Savoie Mont Blanc, CNRS, UMR 5814 74000 Annecy, France}
\affil{$^3$CERN European Organization for Nuclear Research, Esplanade des particules 1, 1211 Geneva, Switzerland}
\affil{$^4$ECAP, Friedrich-Alexander-University Erlangen-Nuremberg, Erwin-Rommel-Str.\ 1, 91058 Erlangen, Germany, \email{kay.graf@fau.de}}
\affil{$^5$Netherlands Institute for Radio Astronomy (ASTRON), Dwingeloo, Netherlands}
\affil{$^6$School of Physical Sciences, The Open University, Walton Hall, Milton Keynes, MK7 6AA, United Kingdom}
\affil{$^7$EGO-European Gravitational Observatory and Scuola Normale Superiore, SNS, Pisa, Italy}


\paperauthor{M. G. Allen}{mark.allen@astro.unistra.fr}{orcid.org/0000-0003-2168-0087}{Observatoire astronomique de Strasbourg, Universit\'{e} de Strasbourg, CNRS, UMR 7550}{}{Strasbourg}{Observatoire Astronomique de Strasbourg}{67000}{France}
\paperauthor{Sample~Author2}{Author2Email@email.edu}{ORCID_Or_Blank}{Author2 Institution}{Author2 Department}{City}{State/Province}{Postal Code}{Country}



  
\begin{abstract}

ESCAPE (European Science Cluster of Astronomy \& Particle physics ESFRI research infrastructures) is an EU H2020 project that addresses the Open Science challenges shared by the astrophysics and and accelerator-based physics and nuclear physics ESFRI projects and landmarks. This project is embedded in the context of the European Open Science Cloud (EOSC) and involves activities to develop a prototype Data Lake and Science Platform, as well as support of an Open Source Software Repository, connection of the Virtual Observatory framework to EOSC, and engaging the public in citizen science. In this poster paper we provide a brief overview of the project and the results presented at ADASS.
  
\end{abstract}

\section{Introduction}
The European Science Cluster of Astronomy and Particle physics research infrastructures (ESCAPE) brings together the astronomy, astroparticle and particle physics communities in a project that is supported by the European Commission Horizon 2020 program call "Connecting ESFRI Infrastructures through Cluster projects" to implement the European Open Science Cloud (EOSC). The ESCAPE project is led by G.~Lamanna (LAPP, CNRS).

ESCAPE brings together the ESFRI projects: the Cherenkov Telescope Array (CTA), the European Solar Telescope (EST), and the cubic-kilometre-sized Neutrino Telescope (KM3NeT); the ESFRI landmark projects including the Facility for Antiproton and Ion Research in Europe (FAIR), the Extremely Large Telescope (ELT), the High Luminosity-Large Hadron Collider (HL-LHC), and the Square Kilometre Array (SKA). Two pan-European International Organizations, the European Organization for Nuclear Research (CERN), and the European Southern Observatory (ESO), are also members of the ESCAPE cluster. The European Virtual Observatory (EURO-VO) is actively engaged in ESCAPE. ESO and other research infrastructures, the European Gravitational-Wave Observatory (EGO-Virgo) and the Joint Institute for VLBI ERIC (JIVE) are also involved.
EOSC is being created with the ambitious goal to be Europe’s virtual environment for researchers to access, manage, exploit and re-use research outputs, open science resources and services for research, innovation and educational purposes. ESCAPE is working to engage researchers from astronomy, astroparticle physics and particle physics in EOSC.

\section{ESCAPE }

The ESCAPE project is organised into five distinct areas as shown in Figure~\ref{ex_fig1} and described in the subsections below.

\articlefigure{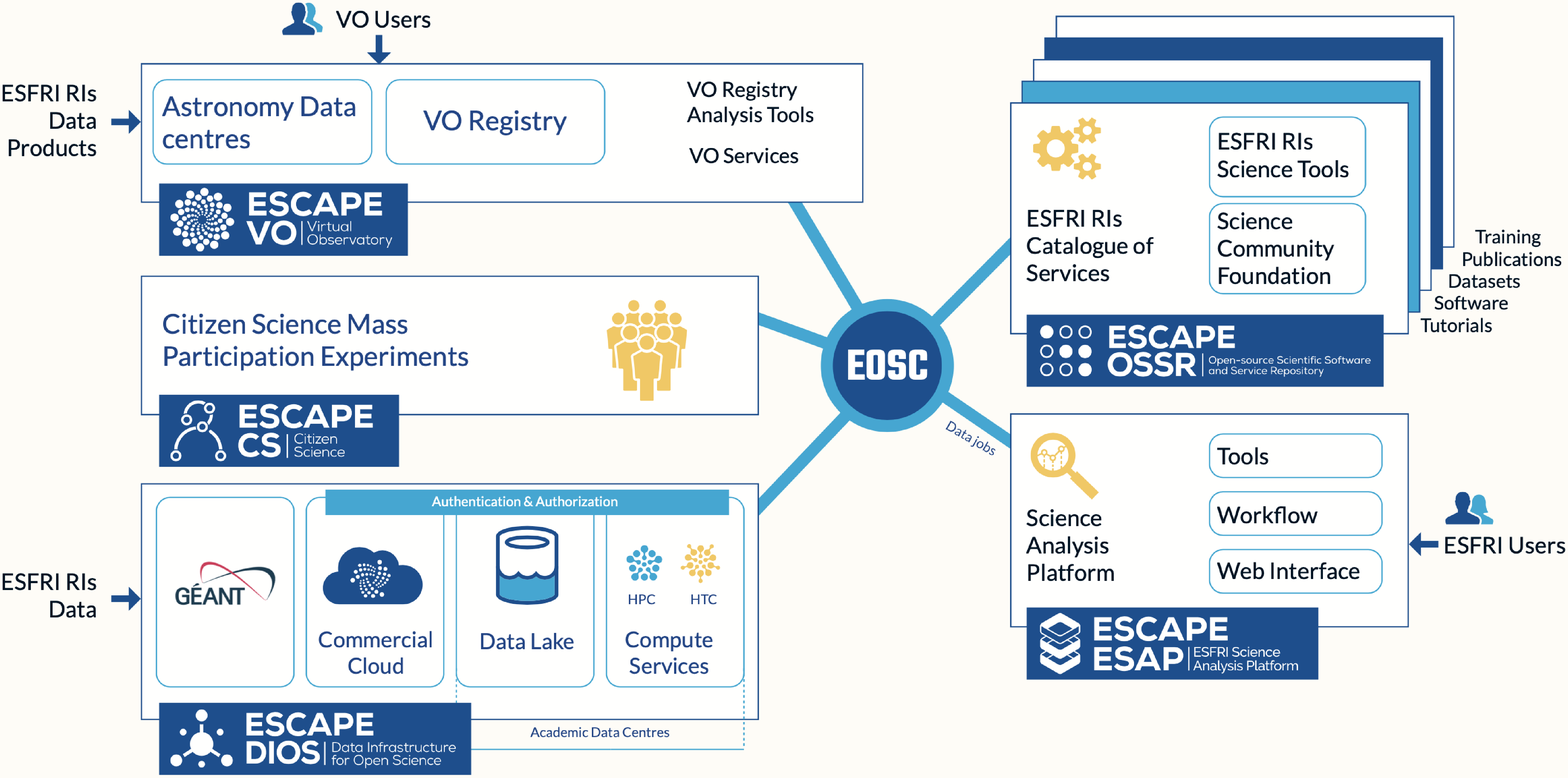}{ex_fig1}{ESCAPE services to be integrated with European Open Science Cloud.}

\subsection{Data Infrastructure for Open Science (DIOS)}

DIOS is a federated data infrastructure of open access data to enable large national research data centres to work together and build a robust cloud-like service to curate and scale up to multi-Exabyte needs.
DIOS is being designed as a flexible and robust data lake in terms of storage, security, safety and transfer, as well as basic orchestration. The Individual physical data stores are being organised into coherent virtualized data infrastructure that manages extremely large data volumes and the user does not necessarily have to know the physical location of the data but accesses it in the cloud via standard interfaces or through science platforms. The interface with heterogeneous computing platforms (HPC, HTC, commercial cloud, etc.) is also implemented.  

\subsection{Open-source Scientific Software and Service Repository  (OSSR)}
The OSSR is an open-access repository to share scientific software and services to the science community and enable open science. It will house astro-particle-physics-related scientific software and services for data processing and analysis, as well as test data sets, user-support documentation, tutorials, presentations and training activities. The goal is to enable a multi-messenger data-driven cooperative approach based on the FAIR principle requirements and will become part of the EOSC global catalogue of services. In a collaborative effort of all ESCAPE partners, common and innovative approaches will be fostered.

\subsection{Virtual Observatory (VO)}
The VO activities of ESCAPE supports the integration of the various astrophysics multi-messenger ESFRI facilities and other research infrastructures into the EOSC through the VO framework. The requirements for the use of the VO are gathered, and activities are pursued in the International Virtual Observatory Alliance (IVOA) for the development of the necessary standards. Training activities are targeted at data providers to implement VO tools and and services, and also toward researchers for scientific use of the data and services via VO tools and science platforms. The connection with EOSC is pursued by identifying the astronomy domain specific data access, discovery and manipulation standards and integrating them with the emerging EOSC systems. Other activitis are designed to add value to astronomy archives via the application of deep learning techniques, and to foster good practices in data stewardship.

\subsection{Science Analysis Platform (SAP)}
The SAP is a platform-service gateway with the capability to access and combine data from multiple collections and stage for subsequent processing and analysis. It allows data discovery and handling of large and distributed data collections. It is a flexible science platform for the analysis of open access data available through EOSC, giving the users the possibility to identify and stage existing data collections for analysis, tap into a wide-range of software tools and packages developed by and in support of the ESFRIs, bring their own custom workflows to the platform, and take advantage of the underlying high performance computing infrastructure to execute those workflows. The ESFRI SAP is tailored to the requirements and the users needs of each of the ESFRI and other research infrastructure members of ESCAPE.

\subsection{Citizen Science (CS)}
The ESCAPE Citizen Science is an ambitious astronomy and astroparticle physics programme to engage a much wider constituency in the FAIR scientific use of the EOSC. The needs of science-inclined public include having a more curated EOSC experience as an entry point, but this does not in itself preclude the public having genuine scientific involvement. 
ESCAPE CS provides the opportunity for the public (and non-specialists in general) to contribute to the implementation of EOSC's crowdsourced scientific data mining, which in turn is designed and supported by the subject-specialist scientific communities. A benefit of ESCAPE CS is public engagement and education, implemented both as bite-sized learning integrated within the CS projects and as public engagement videos. There is also good evidence for CS participants voluntarily extending their scientific literacy through self-guided learning motivated by their CS involvements. In this way ESCAPE CS also supports the next generation of university students, scientists and engineers, who are the future users of the ESFRI facilities.

\section{Early results}

A number of early results have been presented at this ADASS conference. Progress on the OSSR that aims at providing the tools necessary for the communities to share their science products in a harmonised way respecting the FAIR principles is shown in \cite{P3-162_adassxxx}. Another open source software related work is the Hangar data versioning tool geared towards reproducibility and collaboration on numerical datasets \citep{P9-119_adassxxx}. A deep-learning driven, full-event 
reconstruction applied to simulated, IACT events using the CTLearn Python package is shown in \cite{P5-48_adassxxx}. Virtual Observatory related results include the use of the IVOA standards for spatial and temporal coverage, for example the encoding gravitational-wave sky localizations with the Multi Order Coverage data structure in \cite{O4-79_adassxxx}, and also the application to searching for coincidence between LAT/Fermi Exposure Maps and GW Sky Localizations in \cite{P4-252_adassxxx}. The use of IVOA standards for Radio Astronomy visibility data discovery and access is shown in \citep{P9-72_adassxxx} and the status of training events in \citep{P9-49_adassxxx}. In terms of progress on visualisation of astronomy data \citep{O1-68_adassxxx} presents Aladin Lite v3 taking advantage of the GPU with WebGL, and which responds to requests of users, developers and integrators in a context where browser-based applications and science analysis platforms are increasingly important. And progress on the ESAP itself is shown in \cite{P1-59_adassxxx} including how citizen science classification data from the Zooniverse platform can be accessed via ESAP.

\acknowledgements The ESCAPE is funded by the EU Horizon 2020 research and innovation program under the Grant Agreement n.824064.


\bibliography{P3-204}

\begin{thebibliography}{}
\expandafter\ifx\csname natexlab\endcsname\relax\def\natexlab#1{#1}\fi
\expandafter\ifx\csname url\endcsname\relax
  \def\url#1{\texttt{#1}}\fi
\expandafter\ifx\csname urlprefix\endcsname\relax\def\urlprefix{URL }\fi
\providecommand{\eprint}[2][]{\url{#2}}

\bibitem[{{Baumann}(2021)}]{O1-68_adassxxx}
{Baumann}, M. 2021, in ADASS XXX, edited by J.-E. {Ruiz}, \& F.~{Pierfederici}
  (San Francisco: ASP), vol. TBD of ASP Conf. Ser., 999 TBD

\bibitem[{{Berretta}(2021)}]{P4-252_adassxxx}
{Berretta}, A. 2021, in ADASS XXX, edited by J.-E. {Ruiz}, \& F.~{Pierfederici}
  (San Francisco: ASP), vol. TBD of ASP Conf. Ser., 999 TBD

\bibitem[{{Dickinson}(2021)}]{P1-59_adassxxx}
{Dickinson}, H. 2021, in ADASS XXX, edited by J.-E. {Ruiz}, \&
  F.~{Pierfederici} (San Francisco: ASP), vol. TBD of ASP Conf. Ser., 999 TBD

\bibitem[{{Greco}(2021)}]{O4-79_adassxxx}
{Greco}, G. 2021, in ADASS XXX, edited by J.-E. {Ruiz}, \& F.~{Pierfederici}
  (San Francisco: ASP), vol. TBD of ASP Conf. Ser., 999 TBD

\bibitem[{{Louys}(2021)}]{P9-72_adassxxx}
{Louys}, M. 2021, in ADASS XXX, edited by J.-E. {Ruiz}, \& F.~{Pierfederici}
  (San Francisco: ASP), vol. TBD of ASP Conf. Ser., 999 TBD

\bibitem[{{Lutz}(2021)}]{P9-49_adassxxx}
{Lutz}, K. 2021, in ADASS XXX, edited by J.-E. {Ruiz}, \& F.~{Pierfederici}
  (San Francisco: ASP), vol. TBD of ASP Conf. Ser., 999 TBD

\bibitem[{{Nieto Castano}(2021)}]{P5-48_adassxxx}
{Nieto Castano}, D. 2021, in ADASS XXX, edited by J.-E. {Ruiz}, \&
  F.~{Pierfederici} (San Francisco: ASP), vol. TBD of ASP Conf. Ser., 999 TBD

\bibitem[{{Quarenghi}(2021)}]{P9-119_adassxxx}
{Quarenghi}, F. 2021, in ADASS XXX, edited by J.-E. {Ruiz}, \&
  F.~{Pierfederici} (San Francisco: ASP), vol. TBD of ASP Conf. Ser., 999 TBD

\bibitem[{{Vuillaume}(2021)}]{P3-162_adassxxx}
{Vuillaume}, T. 2021, in ADASS XXX, edited by J.-E. {Ruiz}, \&
  F.~{Pierfederici} (San Francisco: ASP), vol. TBD of ASP Conf. Ser., 999 TBD

\end{thebibliography}


\end{document}